\documentclass[a4paper]{jpconf}                                                                       
\usepackage{graphicx}                                                                                 
\usepackage{{amsmath,amssymb}}                                                                        
\usepackage{citesort}                                                                                                
\usepackage{color}

\newcommand{\be}{\begin{equation}}                                                                    
\newcommand{\ee}{\end{equation}}                                                                      
\newcommand{\bea}{\begin{eqnarray}}                                                                   
\newcommand{\eea}{\end{eqnarray}}                                                                     
\newcommand{\bwt}{\begin{widetext}}                                                                   
\newcommand{\ewt}{\end{widetext}}                                                                     

\newcommand{\php}{{\phantom{'}}}

\begin{document}                                                                                      

\vspace*{-3cm}

\title{{\small DIAGRAMMATIC\hspace*{-0.0cm} SELF-CONSISTENT\hspace*{-0.0cm}
    THEORY\hspace*{-0.0cm} OF\hspace*{-0.0cm} ANDERSON\\[-0.3cm] 
    LOCALIZATION FOR THE TIGHT-BINDING MODEL}}

\author{J. Kroha}

\address{{\it Institut f\"ur Theorie der Kondensierten Materie, Universit\"at
Karlsruhe, P.O.B. 6980,
D-7500 Karlsruhe, Fed. Rep. Germany}}

\ead{kroha@physik.uni-bonn.de}

\begin{abstract}
A self-consistent theory of the frequency dependent diffusion coefficient for the Anderson localization problem is presented within the tight-binding model of non-interacting electrons on a lattice with randomly distributed on-site energy levels. The theory uses a diagrammatic expansion in terms of (extended) Bloch states and is found to be equivalent to the expansion in terms of (localized) Wannier states which was derived earlier by Kroha, Kopp and W\"olfle. No adjustable parameters enter the theory. The localization length is calculated in 1, 2 and 3 dimensions as well as the frequency dependent conductivity and the phase diagram of localization in 3 dimensions for various types of disorder distributions. The validity of a universal scaling function of the length dependent conductance derived from this theory is discussed in the strong coupling region. Quantitative agreement with results from numerical diagonalization of finite systems demonstrates that the self-consistent treatment of cooperon contributions is sufficient to explain the phase diagram of localization and suggests that the system may be well described by a one-parameter scaling theory in certain regions of the phase diagram, if one is not too close to the transition point.
\end{abstract}

\section{\label{sec:intro}Introduction}\vspace*{0.2cm}

The concept of localization of a quantum particle in a random potential was introduced by P.W. Anderson [1] in his pioneering work in 1958. Although considerable progress has been made since then, the problem still resists a complete theoretical understanding. In further developing ideas by Thouless [2], Abrahams et al. [3] formulated a real space scaling theory of localization in 1979, which is based on the assumption that the dimensionless conductance $g(L)$ of a sample with finite length $L$ is the only relevant scaling parameter. The assumptions of the one-parameter scaling theory were supported by the work of Wegner [4] and others [5-7], who mapped Anderson's original Hamiltonian onto non-linear $\sigma$ models of interacting matrices and applied perturbative renormalization group calculations in $d = 2 + \varepsilon$ dimensions. Self- consistent theories [8, 9], developed at the same time, also yielded scaling [10] in agreement with these field theoretical treatments. In this way it was established that the Anderson transition is a continuous phase transition characterized by a diverging length, critical exponents and an order parameter.

However, in recent years the validity of the one-parameter scaling theory has been called into question by several new developments: (1) The discovery of large, universal conductance fluctuations in mesoscopic samples [11, 12] has raised the question whether the distribution of conductances rather than the ensemble-averaged conductance $g(L)$ would obey one-parameter scaling [13, 14] or whether more than one scaling parameter would be required in the strong coupling regime [15, 16]. (2) Recent results [17, 18] concerning localization on a Bethe lattice, which were obtained within the supersymmetric matrix model introduced by Efetov [19], show a non-power law singularity at the transition point with an exponentially vanishing diffusion coefficient at the metallic side and a diverging localization length with critical exponent $\nu = 1$ at the insulating side. (3) According to Kravtsov, Lerner and Yudson [20], in the non-linear $\sigma$ model there exist relevant operators from higher gradients, which may indicate a violation of one-parameter scaling. Furthermore, Wegner [21] performed an $\varepsilon$ expansion for the non-linear $\sigma$ model and found in four-loop order a correction to the critical exponent $\nu$ in $d = 3$ dimensions which violates an exact relation for $\nu$ established by Chayes et al. [22]. Thus, it is unclear whether the asymptotic $\varepsilon$ expansion about $d = 2$ can be extended to $3$ dimensions\footnote{See the discussion in ref. [21]}. In this spirit, Zirnbauer [23] has done a non-perturbative calculation in $d = 3$ within Efetov's model using the Migdal-Kadanoff approximate renormalization scheme.

The question of the precise critical behavior near the Anderson transition remains controversial up to now. In the present paper we want to address a different problem, namely how to connect the critical regime with the parame- ters of the Hamiltonian in a quantitative way. To this end, the self-consistent theory originally devised by Vollhardt and W\"olfle [8] is extended to the lattice and fully renormalized in the sense that the calculation of all quantities entering the self-consistency equation is extended to the strong coupling regime. As will be seen below, the self-consistent expansions in terms of small and large disorder, respectively, are equivalent at the level of maximally crossed diagrams. One-parameter scaling is an inescapable consequence of the diagrammatic expansion, unless there exist as yet undiscovered infrared-divergent contributions besides that class of diagrams. However, by comparing with independent numerical renormalization group calculations [24] for the same model one may gain some insight in how wide the critical region around the transition point is in which deviations from one-parameter scaling occur and in which other infrared singularities not contained in the present theory might become important. 

In the next section, quantities which behave non-critical at the Anderson transition are calculated in single-site approximation (CPA). Section 3 contains the solution of the Bethe-Salpeter equation for the density correlation function following Vollhardt and W\"olfle [8], which leads to a frequency dependent diffusion coefficient $D(\omega)$ expressed in terms of the irreducible particle-hole vertex part $U_{{\bf p},{\bf p}'}$. In section 4, $U_{{\bf p},{\bf p}'}$ is calculated by classifying all irreducible particle-hole diagrams in terms of the most important infrared-divergent contributions, and a self-consistent equation is derived in this way. From this equation, the $\beta$ function of the length dependent conductance $g(L)$ is obtained in section 5 along with a discussion of its validity in the strong coupling regime. Some concluding remarks are given in section 6. 

\section{\label{sec:noncritical}Calculation of non-critical quantities}\vspace*{0.2cm}

The model under consideration is defined by the Hamiltonian 
\begin{eqnarray}
H=\sum_{m,n} t_{nm}|n\rangle\langle m| + \sum_{n} V_{n}|n\rangle\langle n| \ ,
\label{eq:hamiltonian}
\end{eqnarray}
with random site energies $V_n$, distributed according to probability $P(V_n)$ and hopping amplitudes $t_{nm}$ from site $m$ to site $n$ ($t_{nn} = 0$). Later evaluations will be done for isotropic nearest neighbor hopping $t_{\langle nm\rangle} = t$ and for box, Gaussian and Lorentzian probability distributions $P_B$, $P_G$ and $P_L$, respectively,
\begin{eqnarray}
P_B(V)&=&\frac{1}{W}  \Theta (W/2-|V|) \ , \nonumber \\
P_G(V)&=&\frac{1}{\sqrt{\pi W^2/6}}  \exp\left(-\frac{V^2}{W^2/6} \right) \ ,
\label{eq:distr} \\
P_L(V)&=&\frac{W}{\pi(V^2+W^2)} \ .   \nonumber 
\end{eqnarray}
Here $\Theta$ is the step function, and the width $W$ plays the role of the disorder parameter. The width of the Gaussian distribution is scaled in such a way that the second moments of $P_B$ and $P_G$ coincide. 

The usual expansion with respect to small disorder $W/t$ is done in terms of Bloch states repeatedly scattering off the impurity levels (extended state expansion). Thus, the single-particle Green function is expanded as
\begin{eqnarray}
\hat G^{R/A}_{nm}(E) &=& \langle 0| (E-H\pm i0)^{-1} |0\rangle \nonumber \\
&=& G^{R/A\,(0)}_{nm} +\sum_{i} G^{R/A\,(0)}_{ni} V_i G^{R/A\,(0)}_{im}  \\ 
&& +\sum_{ij} G^{R/A\,(0)}_{ni} V_i G^{R/A\,(0)}_{ij} V_j G^{R/A\,(0)}_{jm} + \dots\ , \nonumber  
\label{eq:Green_expansion} 
\end{eqnarray}
where
\begin{eqnarray}
G^{R/A\,(0)}_{{\bf p}}(E) = \frac{1}{E-\varepsilon _{\bf p}\pm i0}
\label{eq:Green_free} 
\end{eqnarray}
is the retarded/advanced free Green function of a Bloch state in momentum representation and $\varepsilon_{\bf p} = 2t \sum_{i=1}^{d} \cos p_ia$ is the dispersion on a $d$-dimensional simple cubic lattice with nearest neighbor hopping. 

\begin{figure}[b]
\begin{centering}
\includegraphics[width=0.85\linewidth]{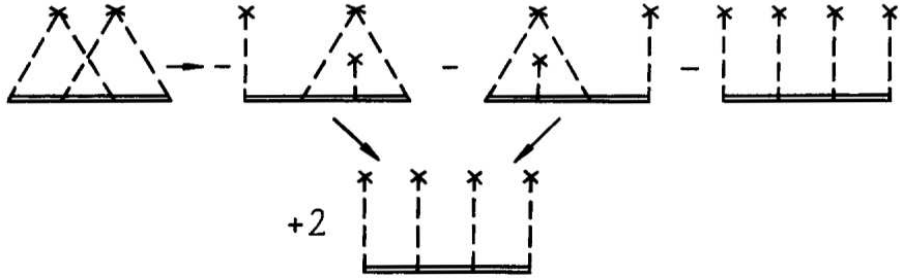}\\[-0.1cm]
\end{centering}
\caption{\label{fig1}
{\small Construction of MOCs: dashed lines represent interactions with the impurity potentials V, horizontal double lines the full Green function G. Each MOC generated from an irreducible diagram (left) corrects for coinciding sites in the respective "direct" diagram with the same topology.}
}
\end{figure}

When the Green function $G$, averaged over the ensemble of all realizations of the random potential, is calculated,
\begin{eqnarray}
G^{R/A}(E) = \langle \hat G^{R/A}(E)  \rangle = \int \prod _n 
\left[ dV_n P(V_n) \right] \hat G^{R/A}(E) \ , 
\label{eq:Green_avg} 
\end{eqnarray}
impurity levels $V$ which happen to be at the same site have to be averaged coherently. This is expressed diagrammatically (fig. 1) by attaching the corresponding impurity lines to the same point and introduces irreducible diagrammatic parts in the calculation of averaged quantities. Then, $G$ satisfies the Dyson equation
\begin{eqnarray}
G = G^{(0)} + G^{(0)} \Sigma G \ ,
\label{eq:Green_Dyson}
\end{eqnarray}
which is easily solved in momentum space:
\begin{eqnarray}
G^{R/A}_{{\bf p}}(E) = \frac{1}{E-\varepsilon _{\bf p} - \Sigma^{R/A}_{\bf p}(E)}  \ .
\label{eq:Green_full} 
\end{eqnarray}
The self-energy $\Sigma_{{\bf p}}^{R/A}(E)$ represents the sum of all irreducible diagram parts with respect to cutting one propagator line $G^{(0)}$. The averaging process causes a technical difficulty which is due to the discreteness of the lattice and,
therefore, is present in any lattice model: In any averaged diagram, lattice sums are restricted to sites which do not occur anywhere else in the diagram, since coinciding sites would change its topological structure. This difficulty is lifted exactly by extending the sums over all lattice sites and correcting for the oversummations by subtracting "multiple occupancy corrections" (MOCs) [25]. MOCs are generated out of any given irreducible diagram by breaking off impurity lines from common sites in all possible ways and subtracting these terms from the original diagram. MOC diagrams may, in turn, require MOCs (next generation) if they have multiple impurity lines at any site. The construction of MOCs is illustrated in fig. 1. 

Alternatively, the Hamiltonian (1) allows an expansion in powers of $t/W$, treating the kinetic term as a perturbation (locator expansion) [25, 26]. The unperturbed on-site Green function, or locator at site $n$, is then given by
\begin{eqnarray}
g^{R/A}_{nm}(E) = \frac{\delta_{nm}}{E-V _{n}\pm i0}  \ .
\label{eq:Locator} 
\end{eqnarray}
Diagrammatically, each term in this expansion is represented by (vertical) locator lines $g$ attached to the corresponding sites and connected by (horizontal) hopping lines $t$. Similarly as in the extended state expansion, it is useful to define a self-resolvent $S(E)$ as the sum of all irreducible diagrams with respect to cutting one hopping line, so that the full, averaged Green function satisfies the Dyson equation
\vspace*{-0.1cm}
\begin{equation}
G = S + S t G \ ,
\label{eq:Locator_Dyson}
\end{equation}
\vspace*{-0.8cm}

\noindent or, in momentum space,
\vspace*{-0.1cm}
\begin{eqnarray}
G^{R/A}_{{\bf p}}(E) = \frac{1}{1/S^{R/A}_{\bf p}-\varepsilon_{\bf p}}  \ .
\label{eq:Green_full_locator} 
\end{eqnarray}
By comparison of eqs. (7) and (10), we can now establish an important relation between the irreducible parts of the weak and strong coupling expansions, respectively [27]:
\begin{eqnarray}
\Sigma^{R/A}_{{\bf p}}(E) = E - [S^{R/A}_{\bf p}(E)]^{-1} \ .
\label{eq:Selfenergies} 
\end{eqnarray}

Let us now turn to the actual evaluation of single-particle quantities. This will be done in single-site or coherent potential approximation (CPA), which amounts to summing up all irreducible diagrams, including MOCs, that do not contain any crossings of lines. The resulting self-consistency equation for the self-energy is given by
\vspace*{-0.4cm}
\begin{eqnarray}
\left< \frac{V- \Sigma^{R/A}_0(E)}{1-[V-\Sigma^{R/A}_0(E)]G^{R/A}_0(E)}
\right>  = 0  \ ,
\label{eq:Selfenergy_avg} 
\end{eqnarray}
where the brackets denote the ensemble averaging and $G_0=(1/N) \sum_{\bf p} G_{\bf p}$. Note that in CPA the self-energy is momentum independent. The CPA can be viewed as the first order of a self-consistent cluster expansion and is known [28] to become exact in the limit of large lattice coordination numbers (or high dimensions). Furthermore, by employing the identity (11) it can be shown [29, 27] that the single-site approximations for the extended state and for the locator expansion are equivalent. Thus, the CPA interpolates between the exact limits of strong and weak disorder. (It cannot account for a correct description of the singularities at the band edges, however.) Therefore, since averaged single-particle quantities are smoothly varying functions at the Anderson transition [30], the CPA is expected to be sufficient for calculating these quantities (fig. 2). 

\begin{figure}[b]
\begin{centering}
\includegraphics[width=0.7\linewidth]{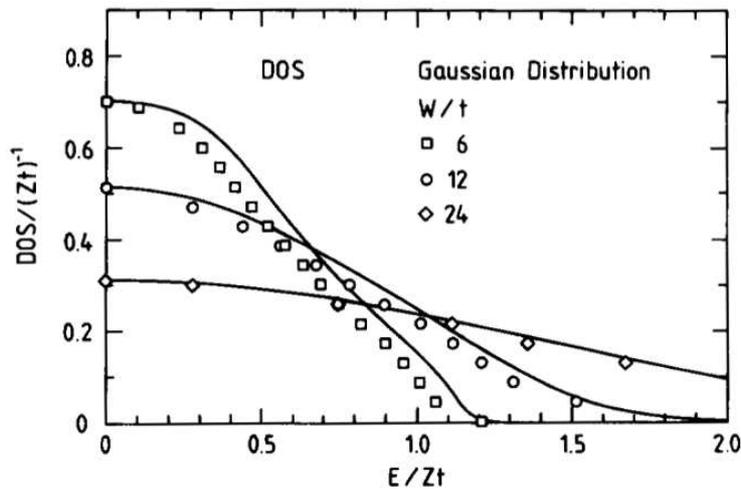}\\[-0.1cm]
\end{centering}
\caption{\label{fig2}
{\small
Disordered density of states in CPA for Gaussian distribution. 
Data points taken from ref. [24] for comparison.}
}
\end{figure}

Kopp [31] formulated a version of the Vollhardt-W\"olfle theory [8] using the locator expansion. While, however, this formulaton was not invariant with respect to a shift of the energy scale, a fully invariant formulation is given in ref. [27]. It will be shown below, that the latter coincides exactly with the formulation using the extended state expansion. Therefore, we will confine ourselves to the extended state formalism in the following.
\vspace*{-0.2cm}

\section{\label{sec:correlation}Correlation functions}\vspace*{0.2cm}

In order to investigate the localized or extended character of states, four- point functions have to be considered, since averaged single-particle functions are translationally invariant. The averaged particle-hole (p-h) Green function $\varPhi_{{\bf p}{\bf p}'}^{RA}(E, \omega, {\bf q}) \equiv \langle \hat G_{{\bf p}_+^{\php}{\bf p}_+'}^R (E + \omega) \hat G_{{\bf p}_-^{\php}{\bf p}_-'}^A(E)\rangle$, with ${\bf p}_{\pm} = {\bf p} \pm {\bf q}/2$ etc., is calculated from the Bethe-Salpeter (B-S) equation
\begin{eqnarray}
\varPhi_{{\bf p}{\bf p}'}^{RA}(E, \omega, {\bf q}) &=& 
G_{{\bf p}_+}^R (E + \omega) G_{{\bf p}_-}^A(E) 
\delta_{{\bf p}{\bf p}'} 
\label{eq:Bethe-Salpeter} \\
&+& G_{{\bf p}_+}^R (E + \omega) G_{{\bf p}_-}^A(E) 
\frac{1}{N} \sum_{{\bf p}''} U^{RA}_{{\bf p}{\bf p}''}(E, \omega, {\bf q})
\varPhi_{{\bf p}''{\bf p}'}^{RA}(E, \omega, {\bf q}) \ , \nonumber
\end{eqnarray}
where $U_{{\bf p}{\bf p}'}^{RA}$, denotes the p-h irreducible vertex function. The B-S equation is supplemented by an exact Ward identity, valid for all frequencies [8]:
\begin{eqnarray}
\Sigma_{{\bf p}_+}^R (E + \omega) - \Sigma_{{\bf p}_-}^A(E) = 
\frac{1}{N} \sum_{{\bf p}'} U^{RA}_{{\bf p}{\bf p}'} 
[ G_{{\bf p}_+}^R (E + \omega) - G_{{\bf p}_-}^A(E) ]
\label{eq:WI}
\end{eqnarray}
Before we will attempt to solve the B-S equation, it is useful to consider the density correlation function 
\begin{figure}[b]
\begin{centering}
\includegraphics[width=0.95\linewidth]{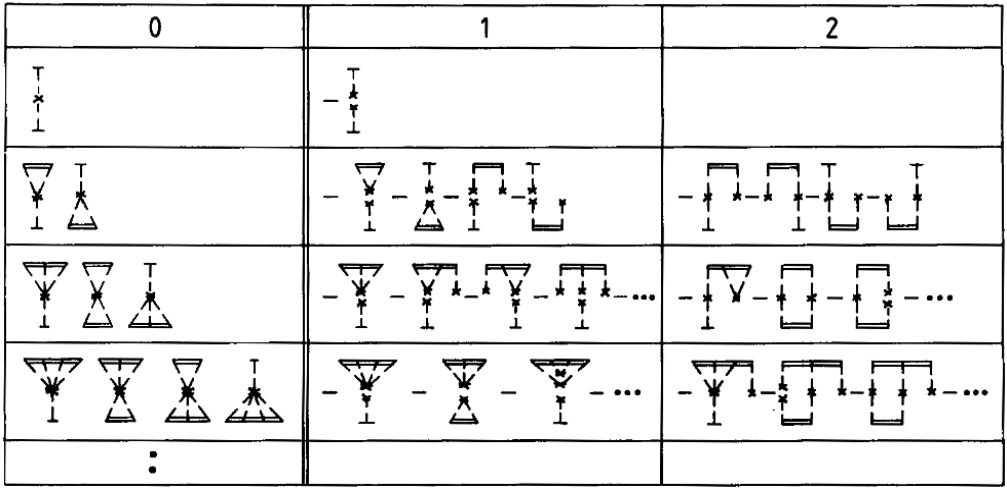}\\[-0.1cm]
\end{centering}
\caption{\label{fig3}
{\small Diagrammatic definition of the single-site vertex $U_0^{RA}$, including MOCs.}
}
\end{figure}
$\varPhi _{\rho\rho}=(1/N^2)\sum _{{\bf p}{\bf p}'}\varPhi _{{\bf p}{\bf p}'}^{RA} ({\bf q})$ in the simplest, i.e. ladder approximation. That is, the irreducible vertex $U$ is calculated as the sum of all irreducible p-h diagrams without crossings (single-site approxmation, fig. 3). Note that this class of diagrams is generated from the CPA diagrams of section 2 via the Ward identity. The result may be obtained by direct
summation (in analogy to the procedure shown in ref. [27]) or by making use 
of the single-site version of the Ward identity, 
$\Sigma _0^R-\Sigma _0^A = U_0^{RA}(G_0^R-G_0^A)$,
\begin{eqnarray}
U_0^{RA} =\frac{\Sigma _0^A(E)-\Sigma _0^R(E+\omega)}{G_0^A(E)-G_0^R(E+\omega)}
\label{eq:SingleSiteVertex}
\end{eqnarray}
The ladder approximation for the total (reducible) vertex is then easily evaluated as (fig. 4)
\begin{eqnarray}
\Gamma_L(\omega, q) = \frac{U_0^{RA}}
{1-U_0^{RA}(1/N)\sum_{\bf p}G_{{\bf p}_+}^RG_{{\bf p}_-}^A}
\stackrel{\omega,q\to 0}{=}
\frac{2\pi iN(E) (U_0^{RA})^2}{\omega +i q^2D_0} \ ,
\label{eq:LadderVertex}
\end{eqnarray}
where $N(E) = (1/\pi) {\rm Im} G_0^A$ is the density of states and
\begin{eqnarray}
D_0=\frac{1}{\pi N(E)}  \frac{1}{N} 
\sum_{\bf p}({\bf  v}_{\bf p}\cdot \hat{\bf q})^2 
({\rm Im}G_{{\bf p}}^A)^2
\label{eq:BareDiffusionconst}
\end{eqnarray}
is the "bare" diffusion coefficient (${\bf v}_{{\bf p}} = \nabla \varepsilon_{{\bf p}}$). For the evaluation of $\Gamma _L$ in the hydrodynamic limit, 
$\omega,\ q \to 0$, we have used the expansion
\begin{eqnarray}
G_{{\bf p}_+}^R(E+\omega)G_{{\bf p}_-}^A(E)  &=&
\frac{1}{U_0^{RA}} \frac{G_{{\bf p}}^A-G_{{\bf p}}^R}{G_{0}^A-G_{0}^R} \nonumber \\
&\times&
\left[
1-\frac{\omega}{U_0^{RA}(G_{0}^A-G_{0}^R)} 
- \frac{1}{2}(G_{{\bf p}}^A-G_{{\bf p}}^R) ({\bf  v}_{\bf p}\cdot \hat{\bf q}) q
\right. 
\label{eq:Expansion} \\
&& \left. - \frac{1}{2}\frac{G_{{\bf p}}^A-G_{{\bf p}}^R}{U_0^{RA}(G_{0}^A-G_{0}^R)}
  ({\bf  v}_{\bf p}\cdot \hat{\bf q})^2 q^2
\right]   +{\cal O} (\omega^2, q^3) \nonumber
\end{eqnarray}
In ladder approximation, the density correlation function reads
\begin{eqnarray}
\varPhi_{\rho\rho}^L =  \frac{1}{N}\sum_{\bf p}G_{{\bf p}_+}^RG_{{\bf p}_-}^A +
\left( \frac{1}{N}\sum_{\bf p}G_{{\bf p}_+}^RG_{{\bf p}_-}^A \right)^2 \Gamma_L 
\stackrel{\omega,q\to 0}{=}
\frac{2\pi iN(E)}{\omega +i q^2D_0} \ .
\label{eq:LadderCorrelationfct}
\end{eqnarray}
The expected diffusion pole structure of (19) is a consequence of the Ward identity (14) and indicates particle number conservation. 
\begin{figure}[t]
\begin{centering}
\includegraphics[width=0.7\linewidth]{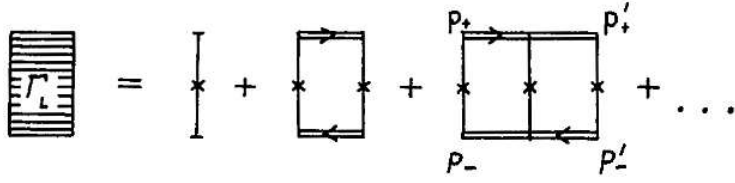}\\[-0.1cm]
\end{centering}
\caption{\label{fig4}
{\small 
Ladder approximation for the total p-h vertex. A full vertical line represents the single-site vertex $U_0$.
}}
\end{figure}
The diffusion coefficient $D$ is expected to vanish at the Anderson transition. A perturbational calculation for $D$, using, e.g., the standard Kubo formula, would therefore require to take into account arbitrarily small contributions, i.e. infinite orders of perturbation theory. This formidable task can, however, be circumvented by observing that a vanishing $D$ is equivalent to a divergence of the density correlation function $\varPhi _{\rho\rho}$ at $\omega\to 0, q\neq 0$. Therefore, for $\omega\to 0$ it is more useful to identify and take into account the {\it largest} infrared divergent contributions to $\varPhi _{\rho\rho}$, which will enter through the irreducible vertex $U_{{\bf p}{\bf p}'}^{RA}$. This will be done in section 4. Here, we first solve the B-S equation in the hydrodynamic limit generally in terms of $U_{{\bf p}{\bf p}'}^{RA}$. 

Eq. (13) can be rewritten in form of a kinetic equation by using
\begin{eqnarray}
G_{{\bf p}_+}^RG_{{\bf p}_-}^A = \frac{G_{{\bf p}_-}^A-G_{{\bf p}_+}^R }
{\frac{1}{G_{{\bf p}_+}^R}-\frac{1}{G_{{\bf p}_-}^A}} \ ,
\nonumber
\label{eq:Relation_bubble}
\end{eqnarray}
multiplying the denominator to the left-hand side and summing over ${\bf p}'$,
\begin{eqnarray}
\Bigl[\omega - ({\bf  v}_{\bf p}\cdot \hat{\bf q})q + \Sigma_{{\bf p}_-}^A \hspace*{-0.1cm}- \Sigma_{{\bf p}_+}^R \Bigr]
\frac{1}{N}\sum_{{\bf p}'} \varPhi_{{\bf p}{\bf p}'}^{RA} 
= (G_{{\bf p}}^A-G_{{\bf p}}^R)
\Bigl( 1\hspace*{-0.05cm} + \hspace*{-0.05cm} \frac{1}{N^2} \sum_{{\bf p}'{\bf p}''} U_{{\bf p}{\bf p}''}^{RA} \varPhi _{{\bf p}''{\bf p}'}^{RA} \Bigr) . \
\label{eq:KineticEquation} 
\end{eqnarray}
Using again the Ward identity (14) and summing over ${\bf p}'$ explicitly incorporates the conservaton of particle number in the form of the continuity equation
\begin{eqnarray}
\omega \varPhi_{\rho\rho} - q \varPhi _{j\rho} = 2\pi i N(E) \ .
\label{eq:ContinuityEquation} 
\end{eqnarray}
It relates $\varPhi _{\rho\rho}$ and $\varPhi _{j\rho} = (1/N^2)\sum _{{\bf p}{\bf p}'}({\bf v}_{{\bf p}}\cdot \hat{\bf q})\,\varPhi _{{\bf p}{\bf p}'}^{RA} ({\bf q})$, 
the density-density and the current-density correlation functions, respectively, to each other. It is this conservation law that governs the hydrodynamic behavior of the system. Therefore, $\varPhi _{{\bf p}} = (1/N)\sum _{{\bf p}}\varPhi _{{\bf p}{\bf p}'}^{RA}$ may be approximated by its projection onto density and current correlations only [31],
\begin{eqnarray}
\varPhi_{\bf p} = A_{\bf p}\varPhi _{\rho\rho} +B_{{\bf p}}({\bf q}) \varPhi _{j\rho} + \Delta \varPhi_{\bf p} \ .
\label{eq:MomentExpansion} 
\end{eqnarray}
Eq. (22) can be viewed as an expansion by moments of the current vertex 
$({\bf v}_{{\bf p}}\cdot \hat{\bf q})$. In fact, contributions 
$\Delta \varPhi _{{\bf p}}$ from higher moments are less divergent and may be neglected in the hydrodynamic limit. Since the critical properties of the
localization transition are contained in the relaxation functions $\varPhi _{\rho\rho}$  and $\varPhi _{j\rho}$\,, the coefficients $A$ and $B$ behave uncritical and may now be obtained from the simple ladder approximation, where 
$\varPhi _{{\bf p}}$, $\varPhi _{\rho\rho}$  and $\varPhi _{j\rho}$ 
are given explicitly,
\begin{eqnarray}
A_{\bf p} &=& \frac{{\rm Im}G_{{\bf p}}^A}{(1/N)\sum_{\bf p}{\rm Im}G_{{\bf p}}^A} \ , 
\nonumber \\
B_{\bf p}({\bf q}) &=& \frac{({\bf  v}_{\bf p}\cdot \hat{\bf q})({\rm Im}G_{{\bf p}}^A)^2}
{(1/N)\sum_{\bf p}({\bf  v}_{\bf p}\cdot \hat{\bf q})^2({\rm Im}G_{{\bf p}}^A)^2} \ . 
\label{eq:Ap-Bp}
\end{eqnarray}
We use eq. (22) to decouple the momentum integrals in the kinetic equation (20) and obtain, after multiplying with $({\bf v}_{{\bf p}}\cdot \hat{\bf q})$ and summing over {\bf p}, a current relaxation equation [8],
\begin{eqnarray}
\Bigl[\omega + 2i {\rm Im}\Sigma_{0}^A + i M(\omega) \Bigr] \varPhi_{j\rho} 
- q \cdot 2 {\rm Im}\Sigma_{0}^A\, D_0\, \varPhi_{\rho\rho} =-q R(E)
\label{eq:Current Relaxation} 
\end{eqnarray}
where the "current relaxation kernel" $M(\omega)$ is given by
\begin{eqnarray}
-i M(\omega) = \frac{2i}{\pi N(E) D_0} \, \frac{1}{N^2} \sum_{{\bf p}{\bf p}'} 
\, ({\bf  v}_{\bf p}\cdot \hat{\bf q})\,  {\rm Im}G_{{\bf p}}^A\,  U_{{\bf p}{\bf p}'}^{RA} 
\, ({\rm Im}G_{{\bf p}'}^A)^2\,  ({\bf  v}_{{\bf p}'}\cdot \hat{\bf q}) 
\label{eq:CurrentKernel}
\end{eqnarray}
and
\begin{eqnarray}
R(E) = \frac{1}{N}  \sum_{{\bf p}} 
\, ({\bf  v}_{\bf p}\cdot \hat{\bf q})^2 \, \frac{1}{2} 
(G_{{\bf p}}^{A\,2} + G_{{\bf p}}^{R\,2} ) \ . 
\label{eq:Rest}
\end{eqnarray}
The density correlation function follows from (21) and (24),
\begin{eqnarray}
\varPhi_{\rho\rho} = \frac{2\pi i N(E) - q^2 
\frac{R(E)}{i\omega + i [1/\tau  + M(\omega)]}}{\omega +i q^2 D(\omega)} \ .
\label{eq:DenstityCorrel_total}
\end{eqnarray}
Here, we have introduced the generalized, frequency dependent diffusion coefficient
\begin{eqnarray}
D(\omega) = \frac{D_0}{1-i\omega\tau + \tau M(\omega)} \ ,
\label{eq:Diffusion_total}
\end{eqnarray}
where $1/\tau = 2 {\rm Im} \Sigma_0^A$ is the single-particle relaxation rate. The ladder diffusion coefficient $D_0$ has been renormalized by a factor essentially given by the {\it inverse} current relaxation kernel, which is in turn expressed in terms of the irreducible vertex $U_{{\bf p}{\bf p}'}^{RA}$. Thus, in the hydrodynamic limit it is indeed sufficient to take into
account the strongest infrared divergencies in $U_{{\bf p}{\bf p}'}^{RA}$, and they will eventually make $D(\omega\to 0)$ vanish at the transition point. Also because of this divergence, the term $\sim q^2R(E)$ in eq. (27) may be dropped for  $\omega,\ q \to 0$.

\begin{figure}[b]
\vspace*{-0.2cm}
\begin{centering}
\includegraphics[width=0.7\linewidth]{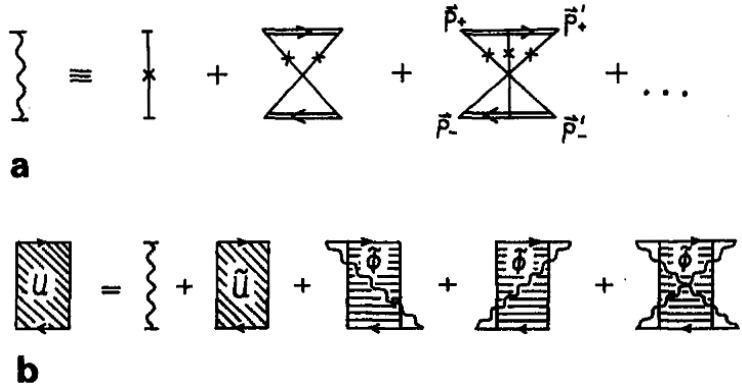}\\[-0.1cm]
\end{centering}
\vspace*{-0.2cm}
\caption{\label{fig5}
{\small
(a) Sum of maximally crossed diagrams $\Lambda _{{\bf p}{\bf p}'}$. (b) Exact classification of all diagrams of the irreducible vertex in terms of cooperons (wavy lines).}
}
\end{figure}

\section{\label{sec:irredvertex}Infrared divergence of the irreducible vertex and self-consistent theory}
\vspace*{0.2cm}

It was shown [8] that the infrared divergence of the diffusion ladder $\Gamma_L$ as an internal part of the total irreducible vertex $U_{{\bf p}{\bf p}'}^{RA}$, is cancelled by vertex corrections• There is, however, another class of diagrams that introduces such a singularity into $U_{{\bf p}{\bf p}'}^{RA}$, namely the class of maximally crossed diagrams known from "weak localization" [32, 33] (fig. 5a). Unlike the diffusion pole, this singularity has no classical analog. It is only present in systems with time reversal symmetry. Due to this symmetry, the ${\bf p} =-{\bf p}'$ singularity ("$2k_F$ singularity", "cooperon") present in the particle-particle channel is carried over to the p-h channel. In this way, the sum of maximally crossed diagrams is obtained as
\begin{eqnarray}
\Lambda_{{\bf p}{\bf p}'}= \Gamma_L({\bf Q}) = 
\frac{2\pi i N(E) (U_0^{RA})^2}{\omega +i ({\bf p}+{\bf p}')^2D_0} \ .
\label{eq:CooperonVertex}
\end{eqnarray}
Here we have introduced ${\bf Q} ={\bf p} + {\bf p}'$.

In any p-h vertex diagram containing a $\Lambda$ block the singularity at ${\bf p} = -{\bf p}'$ will be strongest if $\Lambda$ is crossing the total diagram diagonally, since otherwise the singularity is integrated over and thus weakened. Therefore -- provided that there are no other singular contributions besides cooperons -- the infrared divergent behavior of each diagram contained in the irreducible vertex $U_{{\bf p}{\bf p}'}^{RA}$ can be classified by the number of diagonally crossing cooperons as shown in fig. 5b (Vollhardt and W\"olfle [8, 33]). Wavy lines denote cooperon blocks, while the internal part 
$\tilde \varPhi$ is the sum of all reducible and irreducible p-h diagrams which do not contain any diagonally crossing interaction line. $\tilde U$ represents all irreducible p-h vertex diagrams with the same restriction. 

By flipping over the hole line in the last diagram of fig. 5b and employing time reversal symmetry, the maximally crossing lines $\Lambda_{{\bf p}{\bf p}'}$ are disentangled into ladders $\Gamma _L(Q)$ with the momentum argument ${\bf q}$ replaced by ${\bf Q} ={\bf p} + {\bf p}'$. In this way it is seen that $\tilde \varPhi$ consists of all diagrams contributing to the density correlation function $\varPhi_{\rho\rho}(\omega,Q)$  except those with ladder diagrams at the ends:
\begin{eqnarray}
\tilde\varPhi (Q) &=&\varPhi_{\rho\rho} -R(Q)\,\Gamma_L(Q)\,R(Q) 
- R(Q)\,\Gamma_L(Q)\,\tilde\varPhi(Q) \nonumber \\
&-& \tilde\varPhi(Q)\,\Gamma_L(Q)\, R(Q)
 -  R(Q)\,\Gamma_L(Q)\,\tilde\varPhi(Q)\,\Gamma_L(Q)\, R(Q) \ , \\
&&{\rm with\ } R(q) = \frac{1}{N} \sum_{{\bf p}} G_{{\bf p}_+}^R G_{{\bf p}_-}^A \ ,
\nonumber
\label{eq:Phitilde1}
\end{eqnarray}
or
\begin{eqnarray}
\tilde\varPhi (Q) = \frac{\varPhi_{\rho\rho} - R^2\Gamma_L}{(1+R\Gamma_L)^2} = 
\left( \frac{U_0^{RA}}{\Gamma_L(Q)} \right)^2 [\varPhi_{\rho\rho}(Q) - R^2(Q)\Gamma_L(Q)] \ .
\label{eq:Phitilde2}
\end{eqnarray}
From the last equality it can be seen that the divergencies of the crossing cooperons in the last diagram of fig. 5b are cancelled by the factor $\Gamma_L (Q)^{-2}$ in $\tilde \varPhi(Q)$, leaving the singularity of $\varPhi _{\rho\rho}$ itself. Furthermore, diagrams with less than two diagonally crossing Cooper blocks $\Lambda$ (third--fifth terms in fig. 5b) are less divergent and may be dropped. Thus we have from the second and the sixth diagram:
\begin{eqnarray}
U_{{\bf p}{\bf p}'}^{RA} = \Gamma_L(Q)  + \Gamma_L(Q)\,\tilde\varPhi(Q)\,\Gamma_L(Q) =
(U_0^{RA})^2 \varPhi_{\rho\rho}(Q) \ .
\label{eq:Irred Vertex_total}
\end{eqnarray}
$U_{{\bf p}{\bf p}'}^{RA}$, is governed by the pole structure of $\varPhi_{\rho\rho}$, which in turn feeds back into the diffusion coefficient eq. (28). In this way a self-consistent equation for $D(\omega)$ is obtained:
\begin{eqnarray}
D(\omega)(1-i\omega\tau ) &=& D_0 + \lambda 
\frac{1}{N^2} \sum_{{\bf p}{\bf p}'} 
({\bf v}_{{\bf p}}\cdot \hat{\bf q})\,
\frac{{\rm Im} G_{{\bf p}}^A\, ({\rm Im} G_{{\bf p}'}^A)^2 }{[-i\omega/D(\omega)]+
                                                        ({\bf p}+{\bf p}')^2}\,
({\bf v}_{{\bf p'}}\cdot \hat{\bf q}) 
\nonumber\\
{\rm with\ } \lambda &=& \frac{2{\rm Im} \Sigma_0^A}{[\pi N(E)]^2 D_0} \  . 
\label{eq:D_selfconsistent}
\end{eqnarray}
It is remarkable that this equation is identical to the one derived in the
formulation in terms of a basis of Wannier states [27]: The expansions with respect to small disorder and with respect to small hopping amplitude are equivalent if the effects of quantum interference (cooperons) are taken into account in both formalisms in a self-consistent way. 

Because the cooperon propagator connects momenta of equal size and opposite direction, the term in eq. (33) renormalizing the bare diffusion constant is always negative. In dimensions $d \leq 2$ all states are localized even for arbitrarily small disorder, while for $d>2$ one finds a continuous metal-insulator transition with a scaling relation [8] $\nu = s/(d - 2)$ for the critical exponents $s$, $\nu$ of the conductivity and the localization length, respectively, in agreement with field theoretical treatments [4-6]. 

Eq. (33) has been solved numerically, where the momentum integrals have been evaluated using the isotropic energy-momentum relation that leads to the model bare density of states 
$N_0(E ) = N_d(B/2)^{-1}[1 - E^2/(B/2)^2] ^{-1+d/2}$, i.e. that is defined by 
$N_0(\varepsilon_p)=[S_d/(2\pi)^d]\, p^{d-l}\, |d\varepsilon_p/dp|^{-1}$. Here $B=2Zt$ is the bare band width for lattice coordination number $Z$, $S_d$  is the surface of the $d$-dimensional unit sphere and $N_1 = 1/\pi$, $N_2 = 1/2$, $N_3 = 2/\pi$. In d = 1 this approximation becomes exact. 

In fig. 6 the phase diagrams in the $W$--$E$ plane are shown for box, Gaussian and Lorentzian level distributions $P(V)$. The calculated phase boundaries are in very good agreement with results of exact numerical diagonalization [24]. In particular, the re-entrant behavior for energies outside the bare band is reproduced. This behavior is due to the fact that for energies near the band edge the density of states and thus the diffusion rate first increases with increasing disorder before localization effects start to dominate and drive the system to the insulating phase. The existence of two phase transitions at energies greater than $B/2 = Zt$ can also be seen in fig. 7, where the $\omega = 0$ conductivity is plotted as a function of disorder for various values of the energy $E$.

In fig. 8 the localization length $\xi$, defined by  $\lim_{\omega\to 0}[-i\omega/D(\omega)] = 
1/\xi^2$, is shown in the band center ($E = 0$) as a function of disorder in
$d = 1$, $2$, $3$. The agreement with the numerical results [24] is again very
good for $d = 1$, $2$. For $d = 3$ deviations occur, which may be due to the
fact that an accurate numerical determination of the critical disorder $W_c$
is crucial in this case. Asymptotic power laws in the regions of large and
small localization length are also shown in the figure. In particular, from
eq. (33) it follows that $\xi \sim W^{-1/2}$ for $W\to \infty$. This is the
expected result in a tight binding model because of the following argument: In
the limit of small localization length of a wave function located at site i,
$\xi$ will be determined only by the potential step of order $W$ at the
nearest neighbor sites and will be equal to the decay length of the wave
function into that potential step. For massive particles (which are always
implied in a tight-binding model; \hfill mass m)\hfill \hfill this decay length is 

\begin{figure}[t]
\begin{centering}
\hspace*{3cm}\includegraphics[height=0.98\textheight]{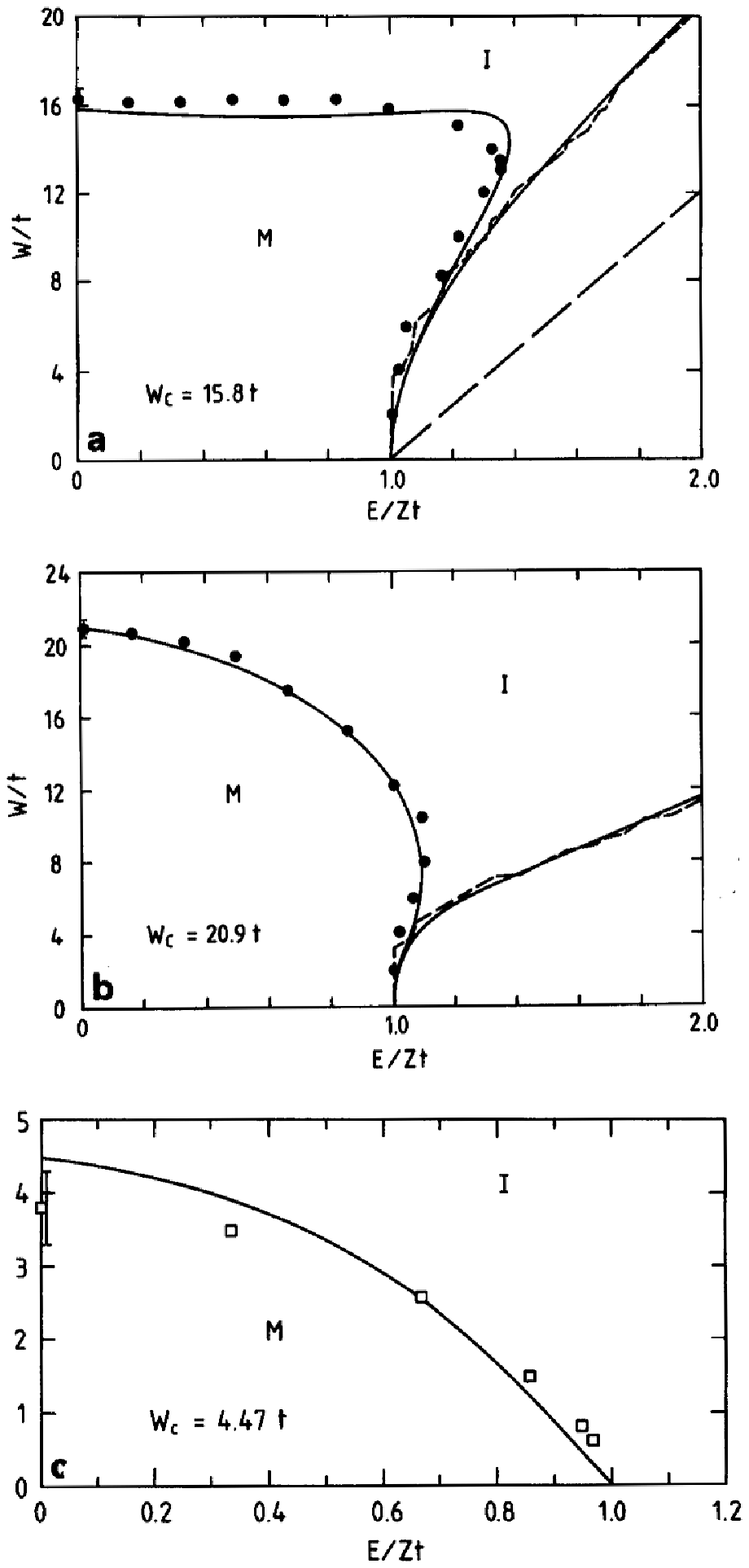}
\vspace*{-2cm}
\end{centering}
{\color{white}
\caption{\label{fig6}}
}
\end{figure}

\noindent given by $1/\sqrt{2m(W-E)}$, thus explaining the 
$W^{-1/2}$ behavior for large $W$. The same result is obtained from a
diagonalization of the corresponding two-site problem in a Wannier basis. As
seen from fig. 8, the crossover from the scaling behavior near the transition
to the atomic limiting behavior indeed occurs at localization lengths smaller
than one lattice constant $a$. Note that the numerical data show a similar
crossover. 

\begin{figure}[t]
\begin{centering}
\includegraphics[width=0.85\linewidth]{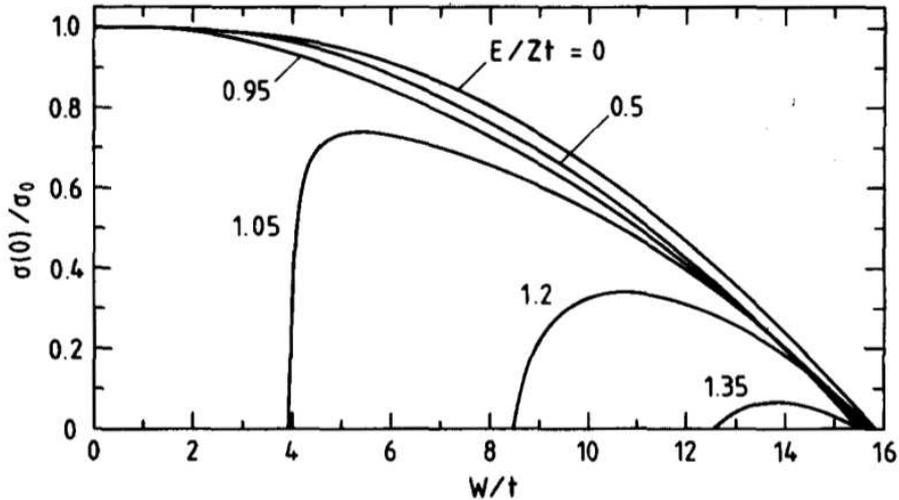}\\[-0.1cm]
\end{centering}
\caption{\label{fig7}
{\small DC conductivity in d = 3 dimensions normalized to the CPA conductivity $\sigma _0 (\omega  = 0)$ as a function of disorder W for various energies E (box distribution).}
}
\end{figure}

In fig. 9 the real part of the dynamical conductivity in three dimensions is shown for 
box-shaped disorder distribution. In the metallic phase ($W < W_c$), ${\rm Re} \sigma(\omega)$ has a dip at $\omega = 0$ with a $\sim \sqrt{\omega}$ singularity, which eventually makes 
${\rm Re} \sigma(0)$ vanish at $W= W_c$ with a singularity $\sim \omega^{1/3}$. In the localized regime ($W > We_c$), a low-frequency behavior ${\rm Re} \sigma(\omega)\sim\omega ^2$ is obtained. For high frequencies $\sigma(\omega)$ merges to the Drude behavior; the conductivity sum rule is fulfilled.

\vfill
\noindent
{\small {\bf $\leftarrow$ Figure 6.} Phase diagram of localization in d = 3 for (a) box, (b) Gaussian, (c) Lorentzian disorder. M: metallic, I: insulating regions. Phase boundary according
to eq. (33) (solid line) and ref. [24] (dots). The error bar of ref. [24] at E = 0 is indicated: $\pm 0.5\,t$. Also shown are the band edge as determined in CPA (solid line in (a), (b)) and numerically in ref. [24] (short-dashed line) and the exact upper bound (long-dashed line in (a)). The critical disorder strength 
$W_c$ for localization in the band center ($E=0$), as obtained from the 
self-consistent theory, is given for each case. 
}

\newpage

\begin{figure}[h]
\begin{centering}
\includegraphics[width=0.65\linewidth]{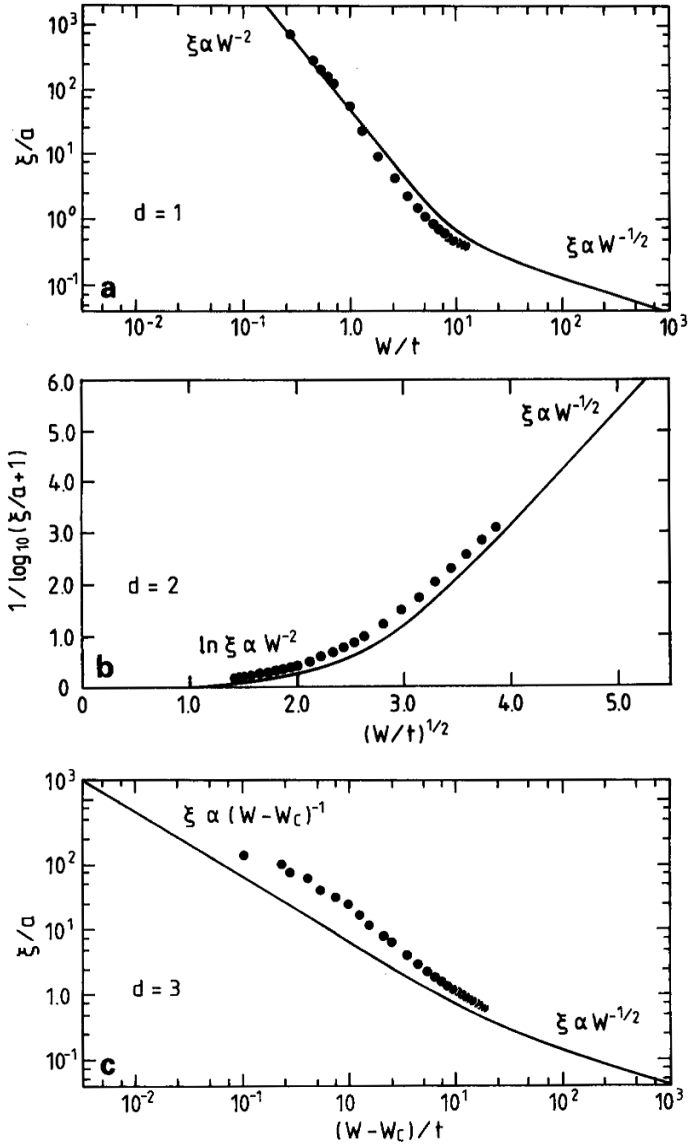}\\[-0.2cm]
\end{centering}
\caption{\label{fig8}
{\small Localization length in units of the lattice spacing as a function of disorder in dimensions d = 1, 2, 3 (a)-(c), for half band filling and box distribution, calculated from eq. (33) (solid lines) and according to ref. [24] (dots). Asymptotic power laws for weak and strong localization are indicated.
}}
\end{figure}

\begin{figure}[t]
\begin{centering}
\includegraphics[width=0.75\linewidth]{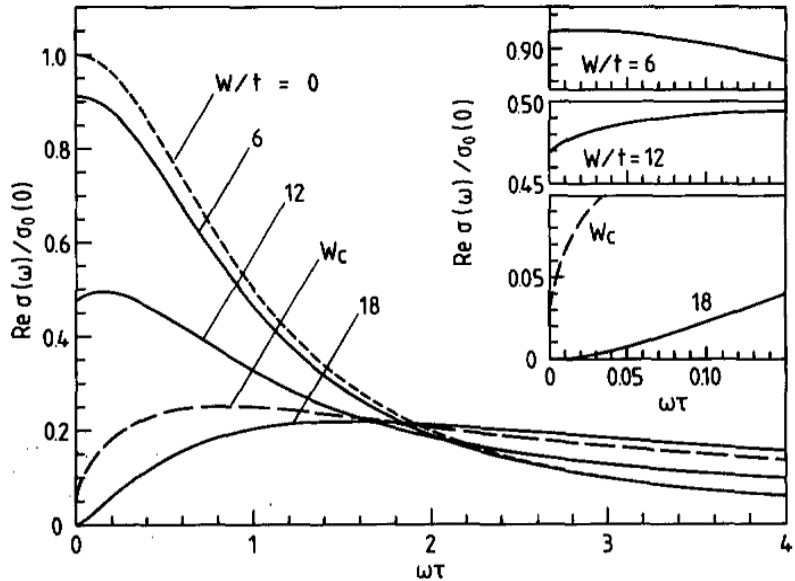}\\[-0.1cm]
\end{centering}
{\small \caption{\label{fig9}
Dynamical conductivity Re $\sigma(\omega)$ normalized to $\sigma_0 (0)$ versus normalized frequency 
$\omega\tau$ for various values of disorder box distribution. Insets show the low-frequency behavior.
}}
\end{figure}

\section{\label{sec:scaling}Scaling function in the strong coupling region}\vspace*{0.2cm}

The controversy between recent results concerning the scaling behavior at the Anderson transition (compare section 1) and the fact that one-parameter scaling is observed in numerical renormalization group calculations [34, 35] at least for energies close to the band center has led the author to re-examine the $\beta$ function for the length dependent conductance derived from the theory presented above. The derivation follows the discussion by Vollhardt and W\"olfle [10]; however, special emphasis is put on the validity of the approximations made in the strong coupling region.\\

{\it Density response in a finite sample}

According to Thouless [2] and Abrahams et al. [3] the dimensionless conductance $g(L)$ of a sample with finite length $L$, $g(L) = (e/\hbar)^2\,L^{d-2}\,\sigma (L)$, should be a relevant scaling parameter. We want to calculate $g(L)$ directly from a density response theory. 

Due to the broken translational invariance in a finite sample, the response is non-local in momentum space, so that the transport coefficient $\sigma (L)$ cannot be calculated in a trivial way from the diffusion coefficient $D(L)$ as in the infinite sample. Instead we consider the density response to an externally applied potential $U(x) = e|{\bf E}|(L - x)$ in position space:
\begin{eqnarray}
\delta \rho (x) = \int_0^L dx' \chi (x - x',\omega = 0)\, U(x') \ .
\end{eqnarray}
Here $\chi (x- x', \omega)$ is the one-dimensional Fourier transform of the 
density response function
\begin{eqnarray}
\chi ({\bf q},\omega ) = \frac{-\omega}{2\pi i}\, \varPhi _{\rho\rho} (q,\omega)
+ N(E) \stackrel{\omega\to 0}{=} 
N(E) \left( \frac{-1}{1+\xi^2q^2} + 1 \right) \ ,
\label{eq:chi}
\end{eqnarray}
with $\xi ~ = \lim_{\omega\to 0} [iD(\omega)/\omega]^{1/2}$, the localization length of the infinite sample. In this way, one obtains without approximations
\begin{eqnarray}
\left. \frac{{\rm d}\rho}{{\rm d}x}\right|_{x=L} = e|{\bf E}|N(E)\,
\frac{1}{2} (1+y)\, {\rm e}^{-y} \ ,
\label{eq:drho_dx}
\end{eqnarray}
where $y = L/\xi$, the sample length scaled to the characteristic length $\xi$, has been introduced. This density gradient at the end of the sample gives rise to a diffusion current $j_{diff} = -e\,D(L)\, {\rm d}\rho/{\rm d}x|_{x= L}$, which in the stationary case is equal and opposite to the electrical current $j_{el}$. \\

{\it Length-dependent diffusion coefficient} 

The effect of a finite sample is that all wave numbers in the system are restricted to values greater than $1/L$. Thus, the length-dependent diffusion coefficient $D(L)$ can be obtained in analogy to eq. (33) as
\begin{eqnarray}
D(L)  = D_0 + \lambda 
\frac{1}{N^2} \sum_{|{\bf Q}|>1/L,\ {\bf p}} 
({\bf v}_{{\bf p}}\cdot \hat{\bf q})\,
\frac{{\rm Im} G_{{\bf p}}^A\, ({\rm Im} G_{{\bf Q}-{\bf p}}^A)^2 }{1/\xi^2 + Q^2}\,
({\bf v}_{{\bf Q}-{\bf p}}\cdot \hat{\bf q})  \ .
\label{eq:DL1}
\end{eqnarray}
The main contribution to the ${\bf p}$ integral in (37) comes from the peak of the imaginary part of the Green function at the Fermi momentum, ${\bf p} = {\bf p}_F$, ${\bf Q}-{\bf p} = {\bf p}_F$. Also, as will be seen below, $D(L)$ can be expressed as an integral over the diffusion pole with $Q$ values restricted to $Q \leq 1/L$. Therefore, the ${\rm p}$ integration in (37) can safely be extended over the complete Brillouin zone, if $p_FL ~> 1$. Furthermore, if the sample size is much larger than the mean free path, the localization length will be the same for the finite sample and for the infinite system. Thus, in eq. (37) the localization length $\xi$ of the infinite sample has been introduced. It is determined in the localized regime by eq. (33):
\begin{eqnarray}
0  = D_0 + \lambda 
\frac{1}{N^2} \sum_{{\bf Q},\ {\bf p}} 
({\bf v}_{{\bf p}}\cdot \hat{\bf q})\,
\frac{{\rm Im} G_{{\bf p}}^A\, ({\rm Im} G_{{\bf Q}-{\bf p}}^A)^2 }{1/\xi^2 + Q^2}\,
({\bf v}_{{\bf Q}-{\bf p}}\cdot \hat{\bf q})  \ .
\label{eq:DL2}
\end{eqnarray}
Subtracting (38) from (37) we obtain
\begin{eqnarray}
D(L)  = - \lambda 
\frac{1}{N^2} \sum_{|{\bf Q}|\leq 1/L,\ {\bf p}} 
({\bf v}_{{\bf p}}\cdot \hat{\bf q})\,
\frac{{\rm Im} G_{{\bf p}}^A\, ({\rm Im} G_{{\bf Q}-{\bf p}}^A)^2 }{1/\xi^2 + Q^2}\,
({\bf v}_{{\bf Q}-{\bf p}}\cdot \hat{\bf q})  \ .
\label{eq:DL3}
\end{eqnarray}
Now the two integrations factorize if $Q \leq 1/L \ll 1/\ell$, where $1/\ell$ is the inverse mean free path measuring the peak width of the Green function. After some algebra one obtains
\begin{eqnarray}
D(L)  = \frac{1}{N(E)}\, \frac{\tilde D_0}{D_0}\, \frac{2}{\pi} S_d
\left( \frac{a}{2\pi} \right)^d\xi^{2-d} \int _0^{1/y} {\rm d}y' 
\frac{y'^{\,d-1}}{1+y'^{\,2}} \ ,
\label{eq:DL4}
\end{eqnarray}
where $\tilde D_0$ is defined by
\begin{eqnarray}
\tilde D_0  = \frac{{\rm Im} \Sigma_0^A}{{\rm Im} G_0^A}\,  
\frac{1}{N} \sum_{{\bf p}} 
({\bf v}_{{\bf p}}\cdot \hat{\bf q})^2\,({\rm Im} G_{{\bf p}}^A)^3 \ .
\label{eq:D0tilde}
\end{eqnarray}
Eqs. (40) and (36) together yielld the dimensionless conductance in the localized regime:
\begin{eqnarray}
g(L)  = \frac{\tilde D_0}{D_0}\, c_d\,  y^{d-2} (1+y)\, {\rm e^{-y}} 
\int _0^{1/y} {\rm d}y' 
\frac{y'^{\,d-1}}{1+y'^{\,2}} \ ,
\label{eq:gL_loc}
\end{eqnarray}
with the dimension-dependent constant $c_d = (1/pi)[S_d/(2~\pi)^d]$. In the metallic regime ($d > 2$) one defines the characteristic length $\xi$ in analogy to the localization length [10], i.e. $\xi \sim (1- \lambda/\lambda_{crit})^{-1/(d-2)}$, and obtains
\begin{eqnarray}
g(L)  = \frac{\tilde D_0}{D_0}\, c_d\, \frac{1}{d-2}\,
[1+\Gamma(d/2)\Gamma(2-d/2)\, y^{d-2}] \ .
\label{eq:gL_deloc}
\end{eqnarray}
Clearly, the quantity $\tilde g(L) \equiv g(L)/(\tilde D_0/D_0)$ depends only on the scaled sample length $y$ and, therefore, obeys a universal scaling law under the conditions posed above. 
The corresponding $\beta$ function is explicitly calculated from eq. (42):
\begin{eqnarray}
\beta (\tilde g)  = \frac{{\rm d} \ln \tilde g}{{\rm d} \ln L} =
d-2 - \frac{c_d}{\tilde g}\, \frac{(1+y)\,{\rm e}^{-y}}{1+y^2} - \frac{y^2}{1+y}\ ,
\label{eq:beta}
\end{eqnarray}
where $y=y(\tilde g)$ from eq. (42).
The function $\beta(\tilde g)$ is shown in fig. 10.

\begin{figure}[t]
\begin{centering}
\includegraphics[width=0.8\linewidth]{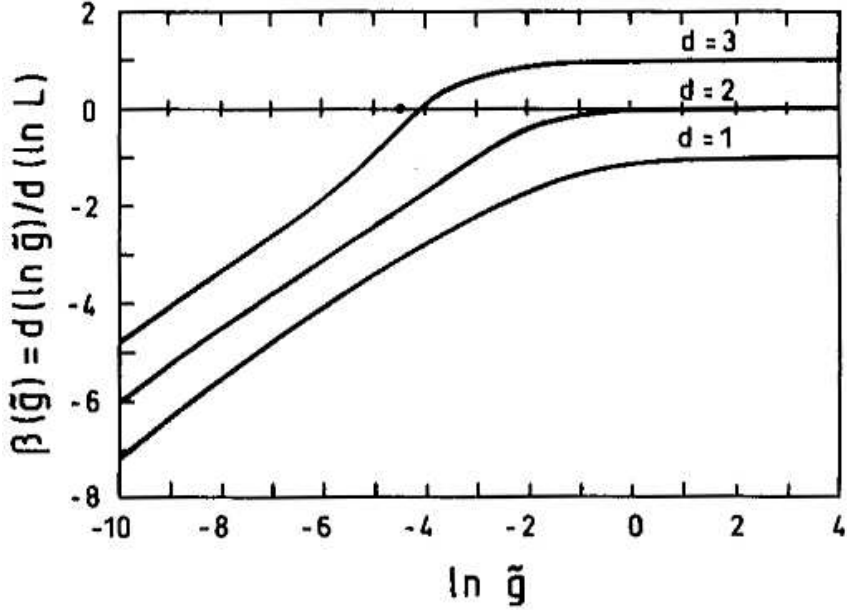}\\[-0.2cm]
\end{centering}
\caption{\label{fig10}
{\small Scaling function $\beta (\tilde g)$ (see text) in d = 1, 2, 3 dimensions. The dot indicates the zero of the
$\beta$ function in d = 3 as determined numerically in ref. [35].
}}
\end{figure}

Summarizing this section, we have found that a one-parameter scaling theory can still be derived from the self-consistent theory in the strong coupling regime. The validity of the one-parameter scaling assumption is however subject to the conditions $p_F L \gg 1$ and 
$L/\ell \gg 1$, which were crucial in the above derivation. They set a lower limit for ~the sample sizes for which scaling can be expected to be observed. In particular, one-parameter scaling must break down at energies close to the band edge, where the Fermi momentum 
$p_F$ approaches $0$. This is in agreement with results of numerical calculations [24, 36]. Furthermore,
we find that only a "renormalized conductance" $\tilde g(L)$ obeys scaling for
large disorder. The renormalization term, i.e., $\ln g(L) - \ln \tilde g(L) =
\ln (\tilde D_0/D_0)$, which is the shift of $\beta (g)$ relative to $\beta (\tilde g)$, has been evaluated numerically. It was found that $\ln (\tilde D_0/D_0)=-0.10$ at the transition point in the band center and that it varies only by a few percent in a range of $W= W_c\pm 4t$ and $E = 0\pm 2t$ 
($d = 3$, box distribution), while it acquires a strong energy dependence near the band edge. Generally, $\ln (\tilde D_0/D_0)$ is a monotonically decreasing function as one moves from the band center to the band edge. The resulting energy dependent shift of $\beta (g)$ is again in agreement with exact numerical d iagonalizations [36]. A physical interpretation of the renormalization factor $\tilde D_0/D_0$ would be required to make contact with the scaling behavior of the actual conductance $g(L)$ in the strong coupling regime, but has not been found yet.

\section{\label{sec:conclusion}Conclusion}
\vspace*{0.2cm}

A self-consistent theory of Anderson localization for the tight-binding model has been presented. The diagrammatic expansions in terms of Bloch and in terms of Wannier states were found to be equivalent, if maximally crossed diagrams are included self-consistently. Considering that no adjustable parame- ters enter the theory, detailed quantitative agreement with results of exact numerical diagonalization of finite systems is found for all quantities available for comparison. The breakdown of the existence of a universal one-parameter scaling function at energies near the disordered band edge, as observed in the numerical calculations, can also be seen in our theory. 

The basic assumptions made in the diagrammatic theory presented here are (1) that at the Anderson transition there be no contributions which are not accessible by perturbation theory and (2) that there be no other infrared divergent diagram classes besides cooperons contributing to the p-h vertex. From the remarkable agreement with numerical data it appears, however, that, if such contributions exist -- as suggested by field theoretical methods --, they should be important in a relatively narrow region around the critical point which is not resolved in our theory nor in the numerical calculations. This would require considering larger and larger system sizes. 

\section*{Acknowledgments}\vspace*{0.2cm}

The author would like to thank Professor Peter W\"olfle for numerous illuminating discussions and for a critical reading of the manuscript. He also acknowledges many useful discussions with Dr. Thilo Kopp.

\end{document}